\documentclass{JHEP3}
\usepackage{graphicx,amsmath,amssymb}
\usepackage{epsfig,multicol}
\usepackage{epsf}
\usepackage{epstopdf}
\input{epsf.sty}

\usepackage{graphicx}
\usepackage{amssymb}
\usepackage{amsmath}

\def\p{\partial}

\def\half{{1\over 2}}
\def\({\left(}
\def\){\right)}
\def\[{\left[}
\def\]{\right]}

\def\e{\begin{equation}}
\def\q{\end{equation}}
\def\m{\begin{eqnarray}}
\def\n{\end{eqnarray}}

\title{Negative spectral index of $f_{NL}$ in the axion-type curvaton model}
\author{Qing-Guo Huang \footnote{huangqg@itp.ac.cn}
\\\small{\em
Kavli Institute for Theoretical Physics China (KITPC), Key Laboratory of Frontiers in Theoretical Physics, Institute of
Theoretical Physics, Chinese Academy of Sciences, Beijing 100190,
China} }

\abstract{
We derive the spectral index of $f_{NL}$ and its running from isocurvature single field and investigate the curvaton models with a negative spectral index of $f_{NL}$ in detail. In particular, a numerical study of the axion-type curvaton model is illustrated, and we find that the spectral index of $f_{NL}$ is negative and its absolute value is maximized around $\sigma_*=\pi f/2$ for the potential $V(\sigma)=m^2f^2(1-\cos{\sigma\over f})$. The spectral index of $f_{NL}$ can be ${\cal O}(-0.1)$ for the axion-type curvaton model. A convincing detection of a positive $n_{f_{NL}}$ will rule out the axion-type curvaton model. In addition, we also give a general discussion about the detectable parameter space for the curvaton model with a polynomial potential.
}


\begin{document}

\section{Introduction}

Curvaton model \cite{Enqvist:2001zp,Lyth:2001nq,Moroi:2001ct} provides an alternative mechanism for generating primordial curvature perturbation.  One distinguishing feature of curvaton model is that it can produce a large local form bispectrum which is proved to be small in the single field inflation model. The size of local form bispectrum is measured by the non-Gaussianity parameter $f_{NL}$. WMAP 7yr data \cite{Komatsu:2010fb} implies $-10<f_{NL}<74$. Recently the curvaton model has been widely studied in the literatures \cite{Dimopoulos:2003az,Chun:2004gx,Sasaki:2006kq,Huang:2008ze,Ichikawa:2008iq,Suyama:2008nt,Li:2008fma,Huang:2008qf,Enqvist:2008gk,Huang:2008rj,Huang:2008bg,Huang:2008zj,Moroi:2008nn,Kawasaki:2008pa,Kawasaki:2008mc,Chingangbam:2009xi,Takahashi:2009dr,Enqvist:2009zf,Takahashi:2009cx,Enqvist:2009eq,Nakayama:2009ce,Enqvist:2009ww,Cai:2010rt,Chingangbam:2010xn,Choi:2010re,Feng:2010tf,Kamada:2010yz}.

Even though the curvaton is a light scalar field compared to the Hubble parameter during inflation, it still slowly rolls down its potential and its dynamics  is expected to have some prints on the observational data. Usually the size of local form bispectrum is considered to be scale independent. However, it is not a generic prediction for the multi-field inflation models, for instance the curvaton model. Recently the authors of \cite{Byrnes:2009pe,Byrnes:2010ft,Byrnes:2010xd} found that the non-linear evolution of curvaton during inflation can induce a detectable scale dependence of $f_{NL}$. Such a scale dependence of $f_{NL}$ is measured by its spectral index $n_{NL}$ which is defined by 
\e
f_{NL}(k)=f_{NL}(k_p)\({k\over k_p}\)^{n_{f_{NL}}},
\q
where $k_p$ is a pivot scale.
Planck \cite{:2006uk} and CMBPol \cite{Baumann:2008aq} are able to provide a 1-$\sigma$ uncertainty on the spectral index of $f_{NL}$ for local form bispectrum as follows, \cite{Sefusatti:2009xu},
\m
\Delta n_{f_{NL}}\simeq 0.1 {50\over f_{NL}}{1\over \sqrt{f_{sky}}}\quad \hbox{for Planck},
\n
and
\m
\Delta n_{f_{NL}}\simeq 0.05 {50\over f_{NL}}{1\over \sqrt{f_{sky}}}\quad \hbox{for CMBPol},
\n
where $f_{sky}$ is the sky fraction. There is no well determined value for the fraction of the sky considered for CMB bispectrum measurements.  For example, in \cite{Komatsu:2010fb}, the authors used for the analysis of non-Gaussianity, the KQ75y7 mask, corresponding to a $f_{sky}$ of $70.6\%$.
Roughly speaking, as long as $| n_{f_{NL}}\cdot f_{NL} |$ is not less than 5 or 2.5, the scale dependence of $f_{NL}$ will be detected by Planck  or CMBPol. The large-scale structure data may provide a more stringent constraint on the scale dependence of $f_{NL}$.
The scale dependence of $f_{NL}$ will become an important observable in the near future.

In this paper we mainly focus on the curvaton model and discuss how it can generate a negative spectral index of $f_{NL}$.
Our paper is organized as follows. In Sec.~2 we consider a general scale dependence of $f_{NL}$ from an isocurvature single field and derive the spectral index and its running of $f_{NL}$. We focus on the axion-type curvaton model where a negative $n_{f_{NL}}$ is predicted in Sec.~3. In Sec.~4 we estimate the detectable parameter space for the curvaton model with a polynomial potential. Some discussions are included in Sec.~5.

\section{The spectral index and running of $f_{NL}$ from an isocurvature scalar field}

In this section we assume that the curvature perturbation is produced by the quantum fluctuation of isocurvature single field $\sigma$ which slowly rolls down its potential at the inflationary epoch. Based on the so-called $\delta N$ formalism \cite{Starobinsky:1986fxa}, the curvature perturbation can be expanded to the non-linear order as follows
\e
\zeta(t_f,{\bf x})=N_{,\sigma}(t_f,t_i)\delta\sigma(t_i,{\bf x})+\half N_{,\sigma\sigma}(t_f,t_i){\delta\sigma}^2(t_i,{\bf x})+... \ , 
\q 
where $N_{,\sigma}$ and $N_{,\sigma\sigma}$ are the first and second order derivatives of the number of e-folds with respect to $\sigma$ respectively. Here $t_f$ denotes a  final uniform energy density hypersurface and $t_i$ labels any spatially flat hypersurface after the horizon exit of a given mode. For simplicity, $t_i$ is set to be $t_*(k)$ which is determined by $k=a(t_*)H_*$ for a given mode with comoving wavenumber $k$. Therefore the amplitude of the
curvature perturbation generated by $\sigma$ takes the form 
\e
P_{\zeta}=N_{,\sigma}^2(t_*)\(H_*\over 2\pi\)^2, 
\label{pnh}
\q 
and the non-Gaussianity parameter $f_{NL}$ is given by 
\e 
f_{NL}={5\over 6}{N_{,\sigma\sigma}(t_*)\over N_{,\sigma}^2(t_*)}. 
\q
Here the horizon crossing approximation is adopted. 
Considering the power spectrum of tensor perturbation $P_T=2H_*^2/\pi^2$, \footnote{We work on the unit of $M_p=1$.} the tensor-to-scalar ratio $r_T$ is defined by
\e
r_T\equiv P_T/P_\zeta={8\over N_{,\sigma}^2(t_*)}. 
\q

In order to work out the scale dependence of $f_{NL}$, let's introduce a new time $t_r (>t_*)$ which is chosen as a time soon after all the modes of interest exit the horizon during inflation. If there is a non-flat potential $V(\sigma)$ for $\sigma$, it will slowly roll down its potential 
\e
3H\dot \sigma\simeq -V'(\sigma),
\q
even though its mass is assumed to be much smaller than the Hubble parameter $H$. The value of $\sigma$ at $t_r$ is related to that at time $t_*$ by
\e
\int_{\sigma_*}^{\sigma_r}{d\sigma\over V'(\sigma)}=-\int_{t_*}^{t_r} {dt\over 3H(t)}.  
\q
Therefore we have
\m
\left. {\p \sigma_r\over \p \sigma_*} \right|_{t_*}&=&{V'(\sigma_r)\over V'(\sigma_*)}, \\
\left. {\p \sigma_r\over \p t_*} \right|_{\sigma_*}&=&{V'(\sigma_r)\over 3H(t_*)}.
\n
One can easily prove 
\e
{d\over d\ln k}F(\sigma_r)={d\over H_*dt_*}F(\sigma_r)=0.
\q
Here we consider ${d\over dt_*}F(\sigma_r)={\p F(\sigma_r)\over \p\sigma_r}(\dot \sigma_* {\p \sigma_r\over \sigma_*}+{\p \sigma_r\over \p t_*})$ and $3H\dot \sigma_*=-V'(\sigma_*)$.
Taking into account that $\sigma_r$ is a function of $\sigma_*$, we have
\m
N_{,\sigma}(t_*)&=&{\p\sigma_r\over \p\sigma_*} {\p N(\sigma_r)\over \p\sigma_r},\\
N_{,\sigma\sigma}(t_*)&=&{\p^2\sigma_r\over \p\sigma_*^2} {\p N(\sigma_r)\over \p\sigma_r}+\({\p\sigma_r\over \p\sigma_*}\)^2  {\p^2 N(\sigma_r)\over \p\sigma_r^2}. 
\n
Since $\p N(\sigma_r)/\p \sigma_r$ and $\p^2 N(\sigma_r)/\p \sigma_r^2$ are scale independent, we find 
\m
{d\ln N_{,\sigma}(t_*)\over d\ln k}&=&\eta_{\sigma\sigma},\\
{d\ln N_{,\sigma\sigma}(t_*)\over d\ln k}&=&2\eta_{\sigma\sigma}+\eta_3 {N_{,\sigma}(t_*)\over N_{,\sigma\sigma}(t_*)},
\n
where the slow-roll equation for $\sigma$ is adopted and 
\m
\eta_{\sigma\sigma}&\equiv& {V''(\sigma_*)\over 3H_*^2},\\
\eta_3&\equiv& {V'''(\sigma_*)\over 3H_*^2}.
\n
From the above results, the spectral index of $P_\zeta$ and $f_{NL}$ are respectively given by
\m
n_s\equiv 1+{d\ln P_\zeta\over d\ln k}=1+2\eta_{\sigma\sigma}-2\epsilon_H,
\n
and
\m
n_{f_{NL}}\equiv {d\ln |f_{NL}|\over d\ln k}=\eta_3 {N_{,\sigma}(t_*)\over N_{,\sigma\sigma}(t_*)},
\n
where
\m
\epsilon_H&\equiv&-{\dot H_*\over H_*^2}.
\n
Our result is the same as that in \cite{Byrnes:2010ft}. The spectral index of $f_{NL}$ is proportional to the third derivative of potential with respect to $\sigma$. A free isocurvature single field cannot generate the scale dependent $f_{NL}$, and the spectral index of $f_{NL}$ is a good quantity to measure the self-interaction of such an isocurvature scalar field.

In addition, the spectral index $n_{f_{NL}}$ may also depend on the scales and its scale-dependence is measured by its running which is introduced in \cite{Byrnes:2010xd}. Here we can easily get the running of the spectral index $n_{f_{NL}}$ generated by an isocurvature field as follows 
\m
\alpha_{f_{NL}}\equiv {dn_{f_{NL}}\over d\ln k}=(2\epsilon_H-\eta_{\sigma\sigma}-\eta_4)n_{f_{NL}}-n_{f_{NL}}^2,
\label{alphafnl}
\n
where
\m
\eta_4\equiv {V'V''''\over 3H^2V'''}. 
\n
Applying the above formula to curvaton model, we get the same result as that in \cite{Byrnes:2010xd}.
Replacing $2\epsilon_H$ by $1-n_s+2\eta_{\sigma\sigma}$, the running of spectral index of $f_{NL}$ becomes
\m
\alpha_{f_{NL}}=(1-n_s+\eta_{\sigma\sigma}-\eta_4-n_{f_{NL}})n_{f_{NL}}.
\n
If $n_{f_{NL}}\sim {\cal O}(1)$, $\alpha_{f_{NL}}$ is roughly the same order of $n_{f_{NL}}$, and then $n_{f_{NL}}$ is not a reliable quantity to measure the scale dependence of $f_{NL}$ any more.
In this paper, we mainly focus on $n_{f_{NL}}$ and don't plan to extend the discussion about $\alpha_{f_{NL}}$ in  detail.

\section{Axion-type curvaton model}

The Nambu-Goldstone boson -- axion -- with shift symmetry $\sigma\rightarrow \sigma+\delta$ is supposed to be a good candidate of curvaton \cite{Lyth:2001nq,Dimopoulos:2003az,Chun:2004gx,Kawasaki:2008mc, Chingangbam:2009xi}. Its potential goes like 
\m
V(\sigma)=m^2f^2\(1-\cos{\sigma\over f}\),
\label{ptaxion}
\n
where $f$ is axion decay constant. If $\sigma\ll f$, the potential is simplified to be
\e
V(\sigma)\simeq \half m^2 \sigma^2+... \ .
\q
It is roughly quadratic and the spectral index of $f_{NL}$ is expected to be small. However, the potential is highly non-quadratic for $\sigma\sim f$ and a large $n_{f_{NL}}$ is expected. 

From the potential in (\ref{ptaxion}), we get
\m
\eta_3=-{\eta_{mm}\over f}\sin{\sigma_*\over f},
\n
where
\e
\eta_{mm}={m^2\over 3H_*^2},
\q
which is related to $\eta_{\sigma\sigma}$ by 
\e
\eta_{\sigma\sigma}=\eta_{mm}\cos(\sigma_*/f).
\label{etass}
\q
We see that $\eta_{\sigma\sigma}=0$ when $\sigma_*=\pi f/2$.
In order to calculate $n_{f_{NL}}$, we need to know $N_{,\sigma}(t_*)$ and $N_{,\sigma\sigma}(t_*)$ as well. It is very difficult to get the analytic formula for these two quantities in the axion-type curvaton model and we will use the numerical method in \cite{Chingangbam:2009xi} to compute them. The numerical calculation is based on the so-called $\delta N$ formalism \cite{Starobinsky:1986fxa} where we start from any initial flat slice at time $t_{ini}$, and then the curvature perturbation will be given by $\zeta(t,{\bf x})=N(t,{\bf x})-N_0(t)$ on the uniform density slicing. Here $N_0(t)$ is the unperturbed amount of expansion.

After inflation, the vacuum energy which governs the dynamics of 
inflation was converted into radiation and the equations of motion of the
universe become
\m
H^2&=&{1\over 3}(\rho_r+\rho_\sigma), \\
\dot \rho_r&+&4H\rho_r=0, \\
\rho_\sigma&=&\half {\dot \sigma}^2+V(\sigma),\\
\ddot\sigma&+&3H\dot \sigma+{dV(\sigma)\over d\sigma}=0, 
\n 
where $\rho_r$ and $\rho_\sigma$ are the energy densities of radiation and
curvaton respectively. It is convenient to introduce some new dimensionless quantities, such as $N(t)=\ln a(t)$, $x=mt$ and 
\e
\theta=\sigma/f.
\q 
Now the Hubble parameter becomes
\e
H=m{dN\over dx}, 
\q 
and the above differential equations of motion are simplified to be 
\m 
{dN\over dx}&=&\[\alpha e^{-4N}+
{f^2\over 3}\(\half({d\theta\over dx})^2+{\tilde V}(\theta)\)\]^\half,\\
{d^2\theta\over dx^2}&+&3{dN\over dx}{d\theta\over dx}+{d{\tilde
V}(\theta)\over d\theta}=0, 
\n 
where 
\e 
V(\sigma)=m^2f^2{\tilde V}(\theta), 
\q 
and 
\e 
\alpha={\rho_{r,ini}\over 3m^2}, 
\q 
and
$\rho_{r,ini}$ is the radiation energy density at $t=t_{ini}$. For
the axion-type curvaton model, 
\e
{\tilde V}(\theta)=1-\cos\theta.
\q
The scale factor can be rescaled to satisfy $a(t_{ini})=1$ and then
$N(t_{ini})=0$. If the inflaton energy decays into radiation very fast, $\alpha\simeq H_{inf}^2/m^2$ which should be much larger than one. Our numerical computation indicates that the numerical result is insensitive to the choice of $\alpha$ for $\alpha>10$. In this paper we set $\alpha=10$. We also adopt sudden decay approximation which says that the curvaton is supposed to suddenly decay into radiation at the time of $H=\Gamma$, or equivalently ${dN/dx}=\Gamma/m$, where $\Gamma$ is the curvaton decay rate. This condition is used to stop the evolution of
curvaton in the numerical calculation. 

\begin{figure}[h]
\begin{center}
\includegraphics[width=13.5cm]{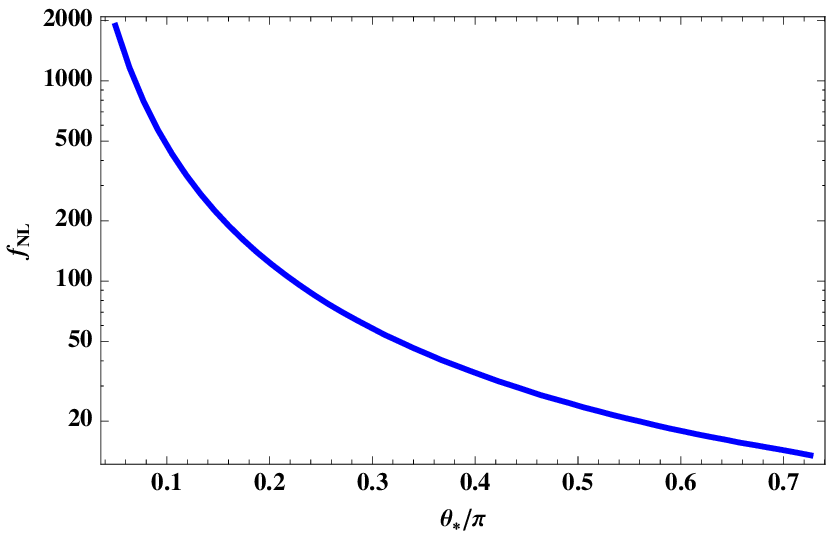}\\
\vspace{.5cm}
\includegraphics[width=13.5cm]{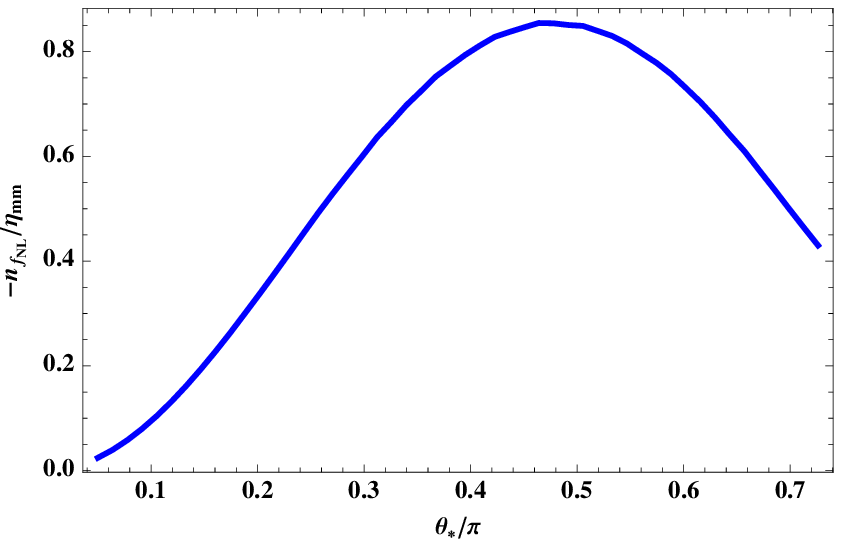}
\end{center}
\caption{$f_{NL}$ and $n_{f_{NL}}$ for axion-type curvaton model. Here we adopt $f=10^{-1}$ and $\Gamma/m=10^{-2}$. }
\label{fig:axion}
\end{figure}

In this section we only plan to illustrate the physics for the axion-type curvaton model and adopt $f=10^{-1}$ and $\Gamma/m=10^{-2}$ in order to shorten the time for numerical calculation. The numerical results of $f_{NL}$ and $n_{f_{NL}}$ are showed in Fig.~\ref{fig:axion}.
A similar plot to the bottom one in Fig.~\ref{fig:axion} is obtained for other choice of the parameters in the model. 
We see that the axion-type curvaton model predicts a negative $n_{f_{NL}}$.  Requiring that the curvaton slowly rolls down its potential during inflation implies $|\eta_{\sigma\sigma}|\ll 1$, but $\eta_{mm}$ can be ${\cal O}(10^{-1}\sim 1)$ around $\theta_*\sim \pi/2$, and the absolute value of $n_{f_{NL}}$ is maximized around $\theta_*= \pi/2$. So $n_{f_{NL}}$ can be ${\cal O}(-10^{-1})$ for axion-type curvaton model. 

If $\sigma_*=\pi f/2$, there is a consistent relation between $n_{f_{NL}}$ and $\alpha_{f_{NL}}$, from Eqs. (\ref{alphafnl}) and (\ref{etass}),
\m
\alpha_{f_{NL}}=(1-n_s-n_{f_{NL}}) n_{f_{NL}}. 
\n
The WMAP 7yr data implies $n_s=0.963$ \cite{Komatsu:2010fb}, and then $\alpha_{f_{NL}}=(0.037-n_{f_{NL}}) n_{f_{NL}}$.


\section{The detectable parameter space for the curvaton model with a polynomial potential}

In this section we consider the curvaton model with a polynomial potential
\m
V(\sigma)=\half m^2\sigma^2+\lambda m^4 \({\sigma\over m}\)^n, 
\n
where the dimensionless coupling $\lambda$ can be positive or negative. For the axion-type curvaton model with $\sigma\ll f$ in Sec.~3, $\lambda=-m^2/(24f^2)$. The curvature perturbation for this kind of curvaton model is discussed in \cite{Enqvist:2008gk,Huang:2008bg,Huang:2008zj,Enqvist:2009zf,Enqvist:2009eq,Enqvist:2009ww,Byrnes:2010xd} in detail. In particular, the authors of \cite{Byrnes:2010xd} only focused on the curvaton model with near quadratic potential and a positive dimensionless coupling $\lambda$. However, in this paper the self-interaction term can be subdominant or dominant. Similar to \cite{Enqvist:2008gk}, we introduce a new parameter 
\m
s\equiv 2 \lambda \({\sigma_*\over m}\)^{n-2}
\n 
to measure the size of the curvaton self-interaction compared to its mass term. 
For the model with negative $s$, $V'(\sigma_*)$ is required to be positive, or equivalently $s>-2/n$; otherwise the curvaton field will run away, not oscillate around $\sigma=0$.
The parameter $\eta_{\sigma\sigma}$  is related to $s$ by
\m
\eta_{\sigma\sigma}={m^2\over 3H_*^2}\[1+{n(n-1)\over 2} s\]. 
\n
Usually $\epsilon_H$ is much smaller than one, and hence a red tilted power spectrum of curvature perturbation cannot be naturally achieved in the curvaton model with quadratic potential \cite{Huang:2008qf}.  Once we take the self-interaction of curvaton into account, this problem is released significantly. 
For $s\in [-{2\over n},-{2\over n(n-1)}]$, 
$\eta_{\sigma\sigma}<0$ and it can make $P_\zeta$ red-tilted.

The spectral index of $f_{NL}$ in this model can be rewritten by
\e
n_{f_{NL}}={5\over 6}{\eta_3\over N_{,\sigma}f_{NL}},
\q
where
\m
\eta_3={n(n-1)(n-2)\over 6}{m^2\over H_*^2}{1\over \sigma_*} s,
\n
for the curvaton model with polynomial potential. 
Combining with the normalization of the curvature perturbation (\ref{pnh}), we obtain
\m
n_{f_{NL}}=\hbox{sign}(N_{,\sigma}) {5n(n-1)(n-2)\over 72\pi \Delta_{\cal R}}{m^2\over H_*^2}{H_*\over \sigma_*}{s\over f_{NL}},
\label{nff}
\n
where $\Delta_{\cal R}=\sqrt{P_\zeta}\simeq 4.96\times 10^{-5}$ \cite{Komatsu:2010fb}. In \cite{Byrnes:2010xd}, the authors did not consider the constraint from the WMAP normalization. We need to stress that the above formula is applicable for any value of $s>-2/n$. Here are three model-dependent free dimensionless  parameters: $m/H_*$, $H_*/\sigma$ and $s$.
Focusing on the curvaton model with near quadratic potential, namely $|s|\ll 1$, $N_{,\sigma}$ is positive and the spectral index of $f_{NL}$ is negative when $s<0$ for a positive $f_{NL}$, such as that in the axion-type curvaton model. For a curvaton model with $n=4$, $\eta_{mm}\sim 10^{-2}$, $H_*/\sigma_*\sim 10^{-1}$ and $|s|\sim 10^{-1}$, $|n_{f_{NL}}\cdot f_{NL}|\sim 6.4$ which is detectable for Planck.

\begin{figure}[h]
\begin{center}
\includegraphics[width=13.5cm]{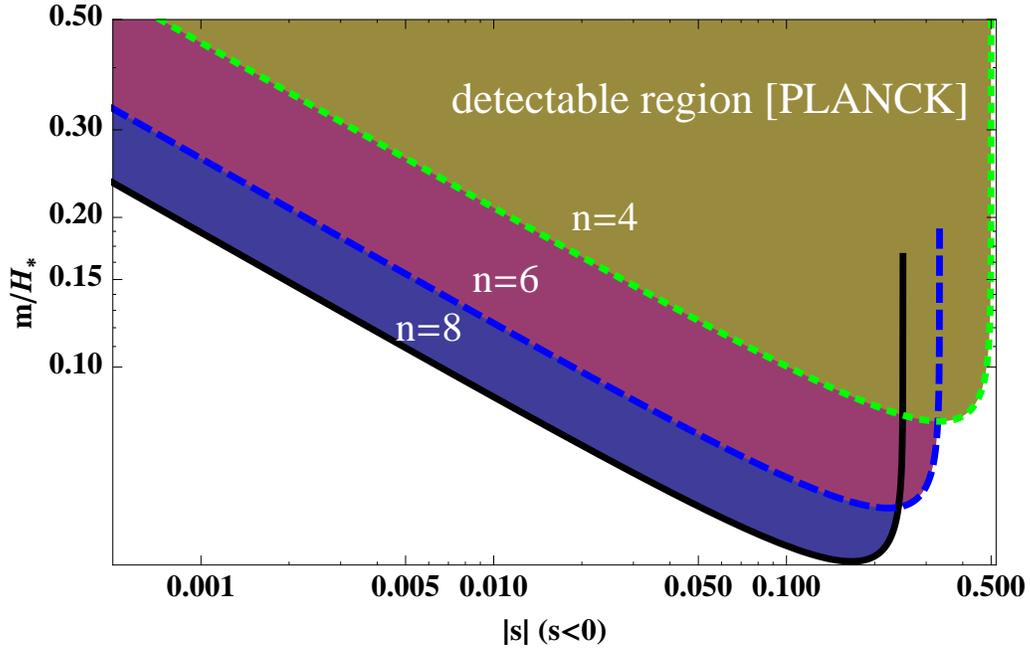}\\
\vspace{.5cm}
\includegraphics[width=13.5cm]{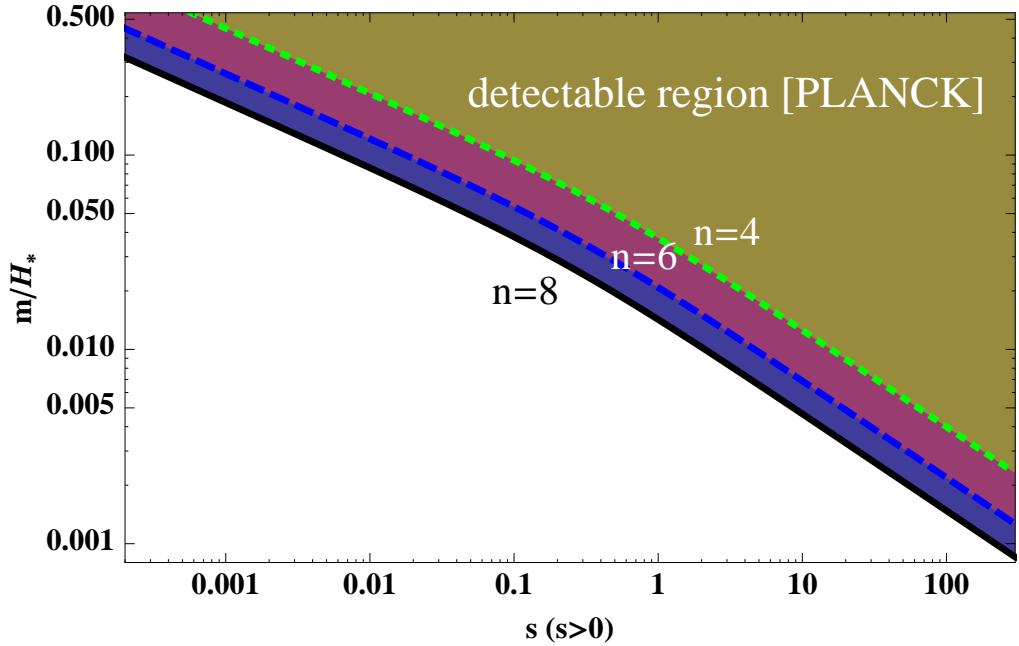}
\end{center}
\caption{The shaded regions illustrate the detectable parameter space for the curvaton model with a polynomial potential.}
\label{fig:poly}
\end{figure}

Here we also want to estimate the typical value of curvaton field during inflation. 
In a long-lived inflationary universe, the long
wavelength modes of the quantum fluctuation of a light scalar field
may play a crucial role in its behavior \cite{Bunch:1978yq}, because its Compton
wavelength is large compared to the Hubble size during inflation. The quantum fluctuation can be taken as a random walk:
\m
\langle \sigma^2 \rangle={H_*^3\over 4\pi^2}t.
\n
On the other hand, the long wavelength modes of the light scalar field are in the slow-roll regime and its dynamics is governed by 
\e
3H_*{d\sigma\over dt}=-m^2\sigma-n\lambda m^3 \({\sigma\over m}\)^{n-1}.
\q
Combining these two considerations, we have
\m
{d\langle \sigma^2\rangle\over dt}={H_*^3\over 4\pi^2}-{2m^2\over 3H_*}\langle \sigma^2\rangle-{2n\lambda m^4\over 3H_*}{\langle \sigma^2\rangle^{n/2}\over m^n}.
\label{qm}
\n
See more discussions in \cite{Huang:2008zj,Bunch:1978yq}.
Typically the vacuum expectation value of $\sigma$ can be estimated as 
\e 
\sigma_*=\sqrt{\langle \sigma^2\rangle}. 
\q 
It is reasonable to suppose that $\sigma_*$ sits at the point where the solution of the differential equation (\ref{qm}) approaches a constant equilibrium value:
\m
\sigma_*=\sqrt{3\over 8\pi^2(1+{n\over 2} s)} {H_*^2\over m},
\n
with $s>-2/n$.
Now Eq.~(\ref{nff}) becomes
\m
n_{f_{NL}}\cdot f_{NL}= \hbox{sign}(N_{,\sigma}) \cdot 2.3\times 10^3  n(n-1)(n-2)({m\over H_*})^3 s\sqrt{1+{n\over 2}s}. 
\n
For the typical choice of the vacuum expectation value of curvaton during inflation, the signal about the scale dependence of $f_{NL}$ depends on the power of $\sigma$ in the self-interaction term, $m/H_*$ and $s$. The detectable parameter space by Planck $(|n_{f_{NL}}\cdot f_{NL}|\gtrsim 5) $ for different $n$ is illustrated in Fig.~\ref{fig:poly}.
The scale dependence of $f_{NL}$ is detectable if the curvaton mass is not too small compared to the Hubble parameter during inflation. If $m/H_*<0.04$, the curvaton self-interaction term must be positive and dominant in order to obtain a detectable scale dependence of $f_{NL}$.

\section{Discussions}

In this paper we discuss the scale dependence of $f_{NL}$ from isocurvature single field and the curvature perturbation can be expanded as
\m
\zeta_k=\zeta_k^L+{5\over 6}f_{NL}(k_p) (1+n_{f_{NL}}\ln{k\over k_p})(\zeta^L *\zeta^L)_k+... \ ,
\n
where $k_p$ is a pivot scale and $*$ denotes a convolution
\e
(\zeta^L*\zeta^L)_k\equiv \int{d^3q\over (2\pi)^3} \zeta^L(q)\zeta^L(k-q). 
\q
The non-linear evolution of isocurvature field induces the scale dependence of 
$f_{NL}$. Similar to the power spectrum of scalar perturbation, one also introduces the so-called spectral index of $f_{NL}$ to measure the tilt of $f_{NL}$. We derive not only the spectral index of $f_{NL}$, but also it running $\alpha_{f_{NL}}$ generated by an isocurvature field. The spectral index $n_{f_{NL}}$ is a good quantity only when it is small compared to one.

Focusing on the curvaton model, we find that the spectral index of $f_{NL}$ can be negative in the axion-type curvaton model (see Fig.~\ref{fig:axion}) and in the curvaton model with small negative self-interaction term. Even though Fig.~\ref{fig:axion} depends on the choice of the parameters, such as $f$ and $\Gamma/m$, in the model, we try some different choices and we find that the similar figures are obtained. If a positive $n_{f_{NL}}$ is supported by the forthcoming cosmological observations, the axion-type curvaton model will be rule out. The spectral index of $f_{NL}$ is a powerful discriminate to distinguishing different curvaton models.

In addition, we also figure out the parameter space for the polynomial potential curvaton model in which such a scale dependence can be detected in the near future. In our consideration, we introduce three parameters, the curvaton mass $m$, the power of curvaton field in the self-interaction term and the relative strength $s$ of self-interaction term compared to the mass term, to characterize the form of curvaton potential. Combining with the estimation of typical vacuum expectation value of curvaton field at inflationary epoch and WMAP normalization, we reach Fig.~\ref{fig:poly} where a large parameter space for detectable scale dependence of $f_{NL}$ is illustrated. We need to stress that our result is applicable even when the self-interaction term becomes dominant.

Finally we also want to point out that the scale-dependent $f_{NL}$ is not a distinguishing feature for the curvaton model. Similar phenomenology can happen in the model \cite{Huang:2009vk} where the entropy fluctuations are converted into the adiabatic perturbation at the end of multi-field inflation due to the geometry of the hypersurface for inflation to end if there are self-interactions for the degrees of freedom along the entropy directions.


\vspace{1.5cm}

\noindent {\bf Acknowledgments}

\vspace{.5cm}

QGH would like to thank P.~Chingangbam for useful discussions. This work is supported by the project of Knowledge Innovation 
Program of Chinese Academy of Science and a grant from NSFC.




\end{document}